# p-(001)NiO/n-(0001)ZnO heterostructures grown by pulsed laser deposition technique

**Bhabani Prasad Sahu, Amandeep Kaur, Simran Arora and Subhabrata Dhar*.**

Department of Physics, Indian Institute of Technology Bombay, Mumbai 400076, India

*Email: dhar@phy.iitb.ac.in

**ABSTRACT:**

NiO/ZnO heterostructures are grown on c-sapphire substrates using pulsed laser deposition (PLD) technique. X-ray diffraction study shows that the ZnO layer epitaxially grows along [0001]-direction on (0001)sapphire surface as expected. While, the epitaxial NiO film is found to be deposited along [001]-direction on the (0001)ZnO surface. Moreover, the presence of three (001)NiO domains laterally rotated by 30° with respect to each other, has also been observed in our NiO films. The study reveals the continuous nature of the NiO film, which also possesses a very smooth surface morphology. In a sharp contrast, ZnO films are found to grow along [0001]-direction when deposited on (111)NiO layers. These films also show columnar morphology. (001)NiO/(0001)ZnO layers exhibit the rectifying current-voltage characteristics that suggests the existence of p-n junction in these devices. However, the behavior could not be observed in (0001)ZnO/(111)NiO heterojunctions. The reason could be the columnar morphology of the ZnO layer. Such a morphology can facilitate the propagation of the metal ions from the contact pads to the underlying NiO layer and suppress the p-n junction effect.

## I. INTRODUCTION:

Semiconductor p–n junctions serve as essential components in many optoelectronic devices including rectifying diodes, transistors, photodetectors, solid-state lasers, light-emitting diodes and solar cells. Certain wide bandgap oxide semiconductors, such as ZnO, $Ga_2O_3$, NiO and $In_2O_3$, have tremendous potential for UV optoelectronics[1–4]. These materials also show exceptional stability in challenging environments and good electrical conductivity along with high optical transparency. Due to the wide and direct bandgap of 3.37 eV and large exciton binding energy of ~60 meV[5], ZnO has emerged as a promising oxide semiconductor for optoelectronic applications. The material often crystallises in wurtzite lattice structure and exhibits n-type conductivity with unintentional background electron concentration of as high as $10^{19} cm^{-3}$, which is often attributed to certain point defects acting as shallow donors[6,7]. Controllable n-type doping can be achieved in ZnO[8,9]. However, reliable and reproducible p-type doping in ZnO poses a significant challenge, primarily due to the high ionization energy of acceptor impurities, limited dopant solubility, and self-compensation effects[10].

Researchers have explored p–n heterojunction devices that combine n-type ZnO with various p-type materials, including GaN, NiO, Si[11–13]. Notably, heterojunctions involving p-GaN/n-ZnO and related structures have received extensive attention, as p-type GaN shares a wurtzite structure and nearly identical lattice constants with ZnO.

Nickel oxide (NiO) is a wide band gap semiconductor with NaCl lattice structure. Its bandgap energy is reported to be ranging from 3.4 to 4.0 eV[14–17]. Most interestingly, stable p-type doping is possible in this material[18,19]. Its electrical conductivity can be adjusted while maintaining high optical transparency. NiO is reported to be grown epitaxially on c-sapphire, c-GaN and MgO substrates, using varied techniques, like pulsed laser deposition[20–24], mist chemical vapor deposition[25], RF magnetron sputtering[26], and molecular beam epitaxy[27] techniques. Among these methods, pulsed laser deposition is the preferred choice for device fabrication due to its superior control over stoichiometry, defect distribution, and the excellent uniformity it provides over a large area of the deposited thin films. Hetero-p-n-junction can thus be grown by integrating p-type NiO with n-type ZnO. Although there are a few reports of the fabrication of such heterojunctions, the critical junction parameters, such as rectification ratio and ideality factor have been observed to be suboptimal and exhibit poor reproducibility. This is primarily attributed to challenges in regulating the crystallite size and film uniformity, leading to diminished electrical and optical properties.

Here ZnO layers are epitaxially grown on c-sapphire substrates using pulsed laser deposition (PLD) technique. The structural, morphological, electrical and luminescence properties of the films are investigated as a function of oxygen pressure during growth. XRD studies show the [0001] directional growth of the ZnO films on c-sapphire. Subsequently, NiO films are grown on (0001) ZnO layers in the same growth chamber using PLD. Interestingly, these NiO films are found to be grown along [001] orientation. In-plane XRD profiles show the existence of three (001)NiO domains rotated laterally by 30° with each other. AFM and SEM studies show the smooth and continuous growth of NiO layers. It has been found that crystal quality of NiO layers deteriorates as the growth temperature is decreased. These (001)NiO/(0001)ZnO heterostructures show rectifying current-voltage characteristics suggesting the existence of p-n junctions. The non-ideal nature of the I-V characteristics can be attributed to the recombination of accelerated electrons and holes in the depletion region.

## II. EXPERIMENTAL DETAILS:

All the samples were deposited on c-sapphire substrate using pulsed laser deposition (PLD) technique. ZnO and NiO pellets are used as targets. The base pressure of the growth chamber was maintained below $1 \times 10^{-5}$ mbar. A KrF excimer laser with wavelength of 248 nm and pulse width of 25 ns was used to ablate the pellet. Energy density of the laser pulse was kept at ∼2 J/cm$^2$ at a frequency of 5 Hz. The substrates were cleaned with trichloroethylene, acetone, IPA and aquas HF before loading into the chamber. First, ZnO samples were grown under various oxygen pressure ($p_{O_2}$) ranging from $1 \times 10^{-4}$ to $1 \times 10^{-1}$ mbar at a fixed growth temperature of 500°C and pulse counts of 10000. Subsequently, NiO films were grown on optimized ZnO/sapphire substrates at different growth temperatures ($T_G$) ranging from 50 to 600°C at an oxygen partial pressure of $2 \times 10^{-2}$ mbar. A portion of ZnO/sapphire template was kept covered by a piece of sapphire during growth. Depositions were carried out for 20000 pulse counts. After the deposition, samples were cooled naturally at the same oxygen pressure used during growth and removed from the chamber at room temperature.

Both in-plane and out-of-plane x-ray diffraction studies were carried out using a Rigaku SmartLab High resolution XRD tool equipped with 9kW rotating Cu anode providing monochromatized Cu $K_\alpha$- radiation. This system has two types of goniometers. $\omega$-$2\theta$ and $\omega$-scans were recorded using a vertical goniometer to investigate the orientation of film in out of plane direction. While the in-plane orientation of the film was examined from $\phi$- and $2\theta_\chi$-$\phi$ scans by a horizontal goniometer. The surface morphology of the samples was studied using atomic force microscopy (AFM) using a Nanoscope Multimode-IV Veeco system and field emission gun scanning electron microscopy (FEG-SEM) using JEOL JSM-7600F setup. Microstructural crystal mapping was studied by Precision electron diffraction (PED) technique using NanoMegas ASTAR® (Brussels, Belgium) with a resolution of ∼1 nm. Sample preparation for PED study was done by Focused ion beam (FIB) lift-out technique using a Helios 5 UC Thermoscientific dual-beam instrument with a spatial resolution of 0.6 nm. For photoluminescence (PL) measurements, the specimens were exposed to a He–Cd laser with a wavelength of 325 nm. The illuminated area of the sample was about 5 mm². The resulting spectra were captured using a 550 mm focal length monochromator equipped with a CCD detector. The current-voltage (I-V) measurements were done using Keithley 6487 Pico ammeter/voltage source using Ni(20 nm)/Au (80 nm) and Ti(20 nm)/Au(80 nm) as contact metal for NiO and ZnO layers, respectively.

## III. RESULTS AND DISCUSSION:

Growth of ZnO layers on c-sapphire substrates are first investigated as a function of oxygen pressure during growth ($p_{O_2}$). XRD studies show the epitaxial nature of all ZnO layers. Further the layers grown at $p_{O_2} = 2 \times 10^{-2}$ mbar has been found to have the lowest screw dislocation density, suggesting the higher crystal quality [see supplementary S.1.2]. The morphology, electrical and luminescence properties of these layers are systematically compared in supplementary S.1. For subsequent growth of NiO layers, ZnO layers grown at $p_{O_2} = 2 \times 10^{-2}$ mbar is used as templates.

A series of NiO layers is grown on (0001)ZnO/c-sapphire substrates at different growth temperatures ($T_G$) ranging from 50 to 600°C. The oxygen pressure during growth is kept fixed at $2 \times 10^{-2}$ mbar. Laser energy density used for ablation is ∼2 J/cm$^2$ in all cases.

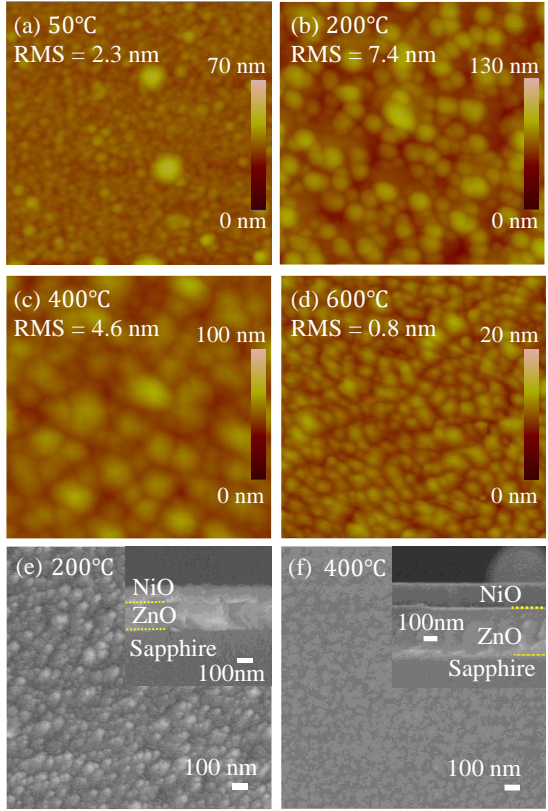

FIG. 1. (a-d) $1\mu m \times 1\mu m$ AFM micrographs of NiO films grown at different temperatures on ZnO/sapphire templates. SEM surface and cross-sectional (inset) image of the NiO films grown at (e) $T_G = 200°C$ and (f) $T_G = 400°C$.

Figure 1(a)−1(d) presents the $1\mu m \times 1\mu m$ AFM micrograph of NiO surfaces grown at different temperatures on ZnO/sapphire templates. The morphology is smooth and continuous in all cases. RMS (root-mean-square) value of surface roughness has been estimated to be as low as 0.8 nm at $T_G = 600°C$. Figure 1(e) and 1(f) shows the surface SEM images of NiO films grown at 200 and 400°C, respectively. The surface appears to have more prominent columnar structures in the sample grown at $T_G = 200°C$ as compared to that grown at $T_G = 400°C$. Cross-sectional SEM images are shown in the respective insets, which show continuous depositions of NiO layers on top of ZnO templates.

Figure 2(a) compares the out-of-plane $\omega$-$2\theta$ XRD scans for all the samples. Peaks corresponding to (002)NiO and (0002)ZnO and their higher order reflections along with sapphire are the only visible features in the scans, suggesting the [001]NiO∥[0001]ZnO∥[0001]sapphire epitaxial relationship in out-of-plane orientation of the films. Note that when NiO is grown directly on c-sapphire and GaN substrates, [111]NiO∥[0001]sapphire/GaN relationship has been found to be maintained[21,28]. One more point to be noticed here is that (002) NiO peak shifts to lower angles [away from (0006) sapphire peak] as the growth temperature is decreased. This suggests an enhancement of tensile strain in the layer along the growth direction with the decrease of the growth temperature. In-plane orientation of the layers grown at $T_G = 400$ and 600°C is investigated by $2\theta_\chi$-$\phi$ XRD scans as shown in fig. 2(b). Only the peaks corresponding to (200)NiO reflections are visible in these scans, which confirms the epitaxial nature of the films. However, it should be mentioned that the layers grown at $T_G < 400°C$ do not show the in-plane epitaxial relationship.

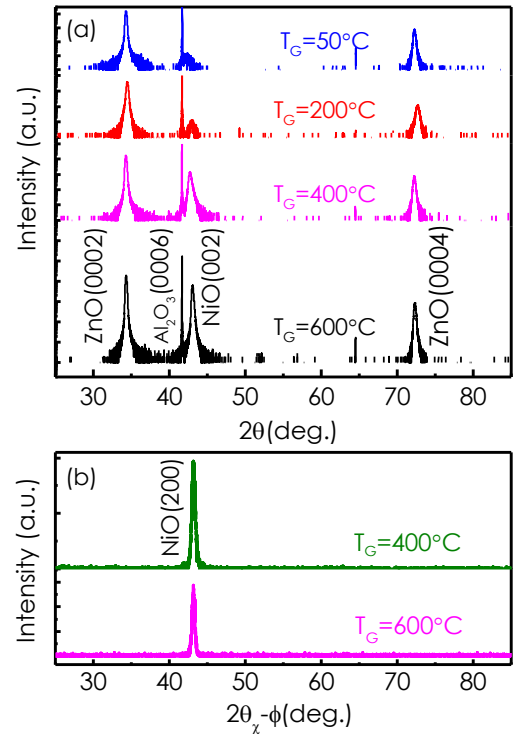

FIG. 2. (a) $\omega$-$2\theta$ and (b) $2\theta_\chi$-$\phi$ XRD scans for NiO layers grown at different growth temperatures.

Figure 3(a) presents the $\phi$ scan profile at (200)NiO reflection for $T_G = 400°C$ grown sample. Twelve equidistant peaks are observed in all samples, suggesting the co-existence of three 30°-laterally rotated cubic domains of NiO when grown on (0001) oriented ZnO films. Figure 3(b) schematically depicts the orientation of three NiO domains on hexagonal (0001) ZnO basal plane. This arrangement corresponds to a lattice mismatch of $(\alpha_{ZnO} - 2a_{NiO})/2a_{NiO} = 2.7\%$. Note that $a_{NiO} = 4.17$ Å and $\alpha_{ZnO} = \sqrt{7}\, a_{ZnO} = 8.58$ Å as shown in Fig. 3 (b).

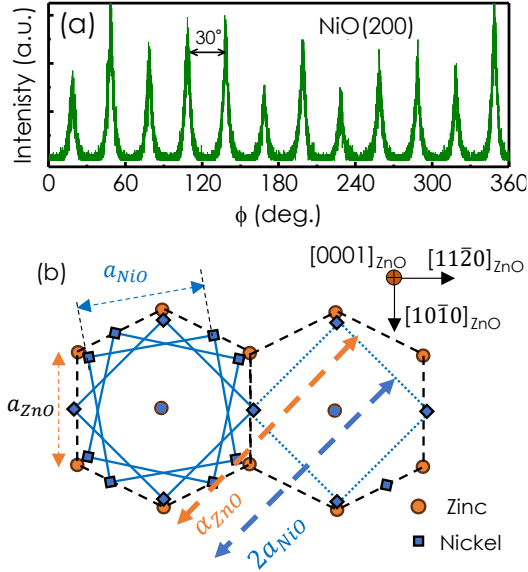

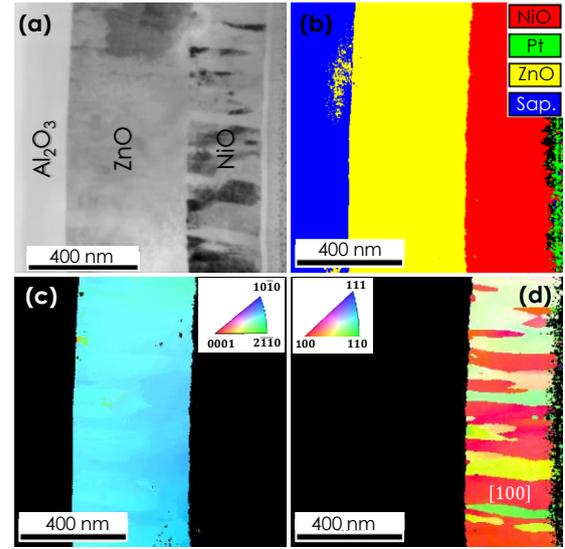

FIG. 3. In-plane wide angle $\phi$-scan at (002) NiO reflection recorded for the film grown at $T_G = 400°C$. (b) Schematic depiction of possible orientations of (100) basal plane of NiO on (0001) ZnO plane.

Figure 4 shows the precision electron diffraction (PED) images obtained by cross-sectional transmission electron microscopy (TEM) on the lamella prepared from the $T_G = 400°C$ grown sample. Figure 4(a) shows the TEM image of the lamella. The growth of a continuous layers of NiO of thickness ~200 nm and ZnO of thickness ~500 nm is quite evident. Figure 4 (b) presents the cross-sectional phase map (EDS map: Zn for ZnO and Ni for NiO) of the lamella. An abrupt interface between the two layers is very clearly visible. Figure 4 (c) and (d) provide the orientation map for the two layers. In case of NiO layer, domains of different orientations could be clearly seen. A thorough investigation reveals that the domains are rotated laterally with respect to each other by angle $n\pi/6$, where $n = 0, \pm1, \pm2, \pm3$. This confirms the observation of Fig. 3.

FIG. 4. (a) Cross-sectional TEM image of the lamella. Precision electron diffraction (PED) images of the lamella: (b) phase map and the orientation maps for (c) ZnO layer and (d) NiO layer in case of the films grown at $T_G = 400°C$.

Densities of screw ($\rho_s$) and edge ($\rho_e$) densities are commonly estimated from the width of the rocking curves ($\Delta\omega$) recorded for planes parallel and perpendicular to the sample surface, respectively, using the following equations:

$$\rho_s = \frac{\Delta\omega_s^2}{4.35 b_s^2}$$
$$\rho_e = \frac{\Delta\omega_e^2}{4.35 b_e^2}$$

where, $\vec{b_s} = <001>$ and $\vec{b_e} = <100>$ are the Burgers vectors for the screw and edge dislocations, respectively[29,30]. Figure 5 presents $\rho_s$ and $\rho_e$ in the NiO layers as a function of $T_G$. It has been found that the value of $\rho_s$ varies between $7 \times 10^{10}$ to $7 \times 10^{11}$ cm$^{-2}$ for samples grown at $T_G \geq 200°C$. However, this value increases to $3 \times 10^{13}$ cm$^{-2}$ for $T_G = 50°C$ sample. On the other hand, the value of $\rho_e$ ranges between $4 \times 10^{11}$ to $3 \times 10^{12}$ cm$^{-2}$ at higher growth temperature.

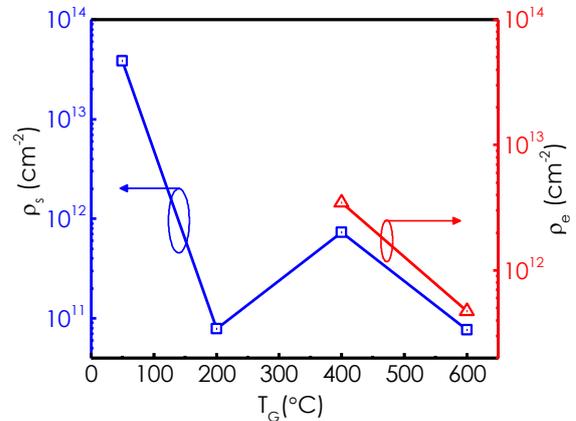

FIG. 5. Screw ($\rho_s$) and edge ($\rho_e$) dislocation densities in NiO layers as a function of $T_G$.

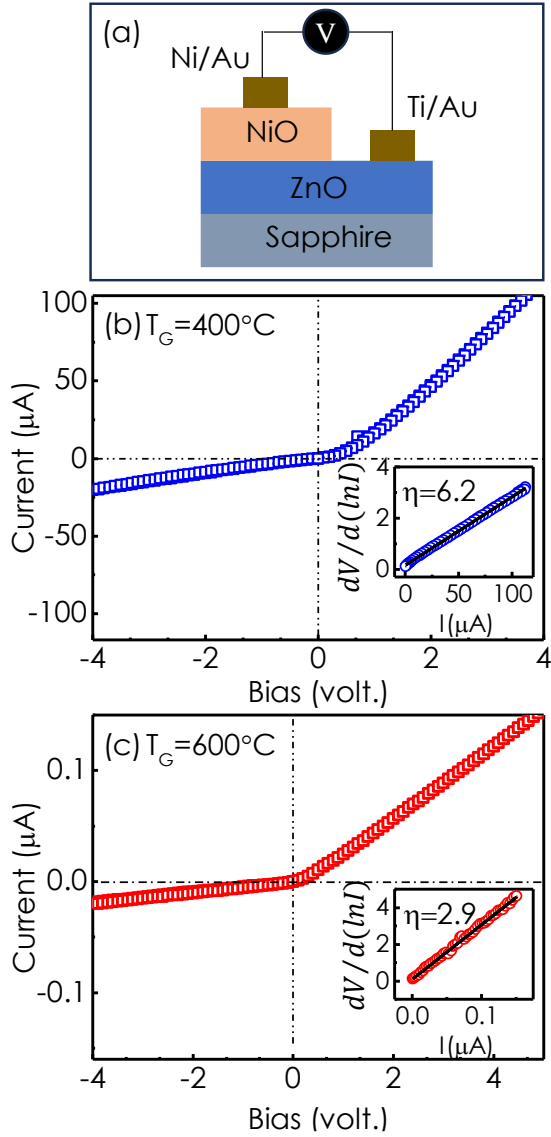

FIG. 6. (a) Schematic diagram of device structure. Current-voltage profiles recorded between Ni/Au and Ti/Au contacts for NiO/ZnO heterostructures grown at (b) $T_G = 400°C$ and (c) $T_G = 600°C$. Inset shows the respective $dV/d(lnI)$ vs $I$ plots.

Electrical measurements are performed on (001)NiO/(0001)ZnO heterojunction devices at room temperature. Ti(20 nm)/Au(80 nm) and Ni(20 nm)/Au(80 nm) contact pads are fabricated on ZnO and NiO sides of these devices, respectively. Figure 6(a) depicts the schematic diagram of the device structure . Ohmic nature of the Ni/Au contact pads on NiO layers and Ti/Au contact pads on ZnO layers are clearly evident [see supplementary S2.1(a)]. Figure 6 (b) and (c) present the room temperature current-voltage profiles recorded between Ni/Au and Ti/Au contacts of NiO/ZnO heterostructures grown at $T_G = 400°C$ and 600°C, respectively. The profiles clearly display rectifying behavior, suggesting the existence of p-n junctions in these devices. It has to be noted that (001)NiO/(0001)ZnO heterojunction devices with NiO layer grown at lower temperatures do not show rectifying characteristic. This might be due to the degradation of crystalline quality and surface morphology of the NiO film observed at low growth temperatures as verified by XRD and SEM studies. In a p-n junction diode, I-V characteristics can be expressed as[31];

$$I = I_s \left[ exp\left(\frac{q(V - IR_s)}{\eta k_B T}\right) - 1 \right]$$

where, $I_s$ the reverse saturation current flowing through the depletion region, $V$ the applied bias, $R_s$ the series resistance, $\eta$ the ideality factor, T the measurement temperature, $q$ the electronic charge and $k_B$ the Boltzmann constant. The above equation can be rewritten as[31]

$$\frac{dV}{d(lnI)} = IR_s + \frac{\eta k_B T}{q}$$

The data shown in figure 6(b) and 6(c) are plotted as $dV/d(lnI)$ vs $I$ in the respective insets for the sample grown at $T_G = 400°C$ and $T_G = 600°C$. Evidently both the plots show linear behavior. The ideality factor (η) is obtained from the intercepts as 6.2 for $T_G = 400°C$ and 2.9 for $T_G = 600°C$. Both the values are quite high even though they are comparable (or lower than) with those reported earlier for p-NiO/n-ZnO heterojunctions[11,32–34]. This high η value has been attributed to the accelerated recombination of electrons and holes in depletion region [35].

It should be noted that we have also grown ZnO layers on top of (111)NiO/Sapphire templates. The SEM image of the as-grown ZnO films show the columnar surface morphology [see supplementary S3.1(a)]. The [0001] orientation of ZnO layers on [111]NiO layers has been verified by out-of-plane $\omega$-$2\theta$ XRD scans [see supplementary S3.1(b)]. Interestingly, the heterojunction does not show any rectifying behaviour, which may be due to the columnar nature of the ZnO film. It is plausible that the metal layers deposited on ZnO surface can propagate through the pores and make contacts with the underlying NiO layer.

## IV. CONCLUSION:

NiO layers can be epitaxially grown along [001] direction on c-ZnO films. This is in contrast with the growth of NiO on c-sapphire and c-GaN substrates, where [111] orientated NiO films are reported to grow. X-ray diffraction studies as well as electron diffraction investigations demonstrate the presence of three (001)NiO domains rotated laterally by 30° with respect to each other, on (0001)ZnO plane. AFM and SEM studies confirm the continuous nature of the film and a very smooth surface morphology. The surface roughness is found to be as low as 0.8 nm. In contrast, ZnO films are found to grow in [0001] orientation on (111)NiO layers. These films show columnar morphology. The current-voltage (I-V) characteristics of (001)NiO/(0001)ZnO layers show the rectifying behavior, suggesting the existence of p-n junction in these devices. However, no rectifying behavior has been observed in case of (0001)ZnO/(111)NiO devices. The columnar morphology of the ZnO layer may help in propagating the metal from the contact pads down to the underlying NiO layer, which could be the reason for not observing any rectifying behavior in these devices.

**Acknowledgement**

We acknowledge financial support provided by Science and Engineering Research Board (SERB) of Government of India (Grant No.CRG/2022/00l852). We would like to thank Industrial Research and Consultancy Centre (IRCC) and Centre of Excellence in Nanoelectronics (CEN), IIT Bombay for using various facilities. The authors would like to thank Sana Ayyubi for her assistance in wire bonding of contact pads.

Supplemental material for

# p-(001)NiO/n-(0001)ZnO heterostructures grown by pulsed laser deposition technique

**Bhabani Prasad Sahu, Amandeep Kaur, Simran Arora and Subhabrata Dhar[*].**

Department of Physics, Indian Institute of Technology Bombay, Mumbai 400076, India

*Email: dhar@phy.iitb.ac.in*

# S1. Growth of ZnO on c-sapphire

A set of ZnO layers are grown at different oxygen pressures ($p_{O_2}$) ranging from $1 \times 10^{-4}$ to $1 \times 10^{-1}$ mbar for a fixed growth temperature of 500°C.

Morphology

Figure S1.1(a-d) represents the AFM surface images of ZnO sample grown on c-sapphire at different oxygen pressure. A uniform surface morphology of the films has been observed. The RMS surface roughness clearly decreases with the oxygen pressure in these samples. Figure S1.1(e) shows the SEM surface image of sample grown at $p_{O_2} = 2 \times 10^{-2}$ mbar. Inset shows the cross-sectional image of the same sample. The film clearly shows smooth surface morphology and good continuity. The film thickness in all cases is found to be ~600 nm.

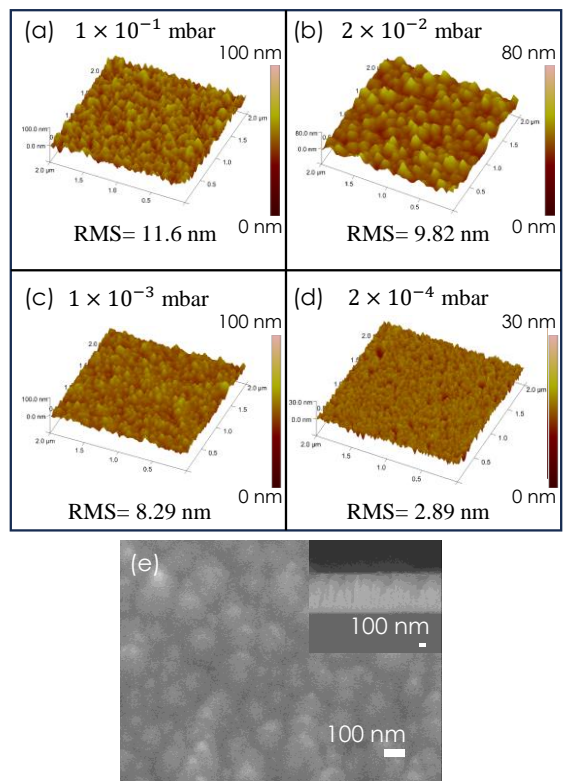

FIG. S1.1. (a-d) AFM micrographs of ZnO surfaces at different oxygen pressures. (e) Surface and cross-sectional (inset) SEM images of the film grown at oxygen pressure of $2 \times 10^{-2}$ mbar.

Crystallinity:

Figure S1.2(a) presents the high resolution $\omega$-$2\theta$ XRD scans for ZnO films grown on c-sapphire substrates at different oxygen pressures. In all the cases, (0002) and its higher order reflection (0004) are the only visible features associated with the wurtzite phase of ZnO. This shows [0001] directional growth of ZnO layers on c-sapphire substrates. Figure S1.2(b) shows the $\omega$ - rocking curve of (0002)ZnO reflection for the sample grown at $p_{O_2} = 2 \times 10^{-2}$ mbar. Figure S1.2(c) presents the FWHM of (0002)ZnO rocking curve as a function of oxygen pressure during growth. It has to be noted that the minimum broadening has been observed in the sample grown at $p_{O_2} = 2 \times 10^{-2}$ mbar.

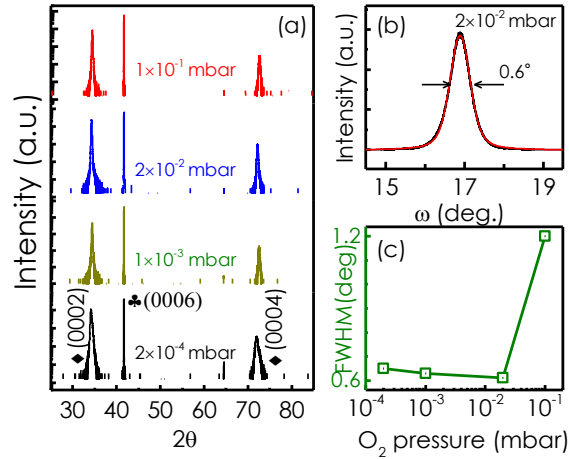

FIG. S1.2 (a) $\omega$-$2\theta$ XRD scan for all samples grown at various oxygen pressures. Symbols ♦ and ♣ represents peaks from ZnO and sapphire, respectively. (b) $\omega$- rocking curve of (0002)ZnO reflection for the film grown at $2 \times 10^{-2}$ mbar. (c) FWHM of the(0002)ZnO rocking curve as a function of oxygen pressure.

Figure S1.3(a) represents the $2\theta_\chi$-$\phi$ XRD profile for samples grown at different pressures. $(10\bar{1}0)$ and higher order reflections from ZnO are observed at relatively lower oxygen pressures. It should be noted that $(11\bar{2}0)$ reflection of ZnO starts appearing when oxygen pressure is more than $1 \times 10^{-3}$ mbar. This suggests the inclusion of grains with other orientations when growth takes place at high enough oxygen pressure. Figure S1.3(b) shows wide angle $\phi$ - scan profiles recorded for $(10\bar{1}0)$ ZnO reflection for these samples. Baring the sample grown at $p_{O_2} = 1 \times 10^{-1}$ mbar, all other samples show six-equidistant peaks that confirms the six-fold symmetry of the $(10\bar{1}0)$ plane of ZnO. All these observations clearly demonstrate the epitaxial nature of the ZnO layers. More than six peaks are evident in $1 \times 10^{-1}$ mbar sample, which can be attributed to the inclusion of misoriented grains in the layer as observed in figure S1.3(a).

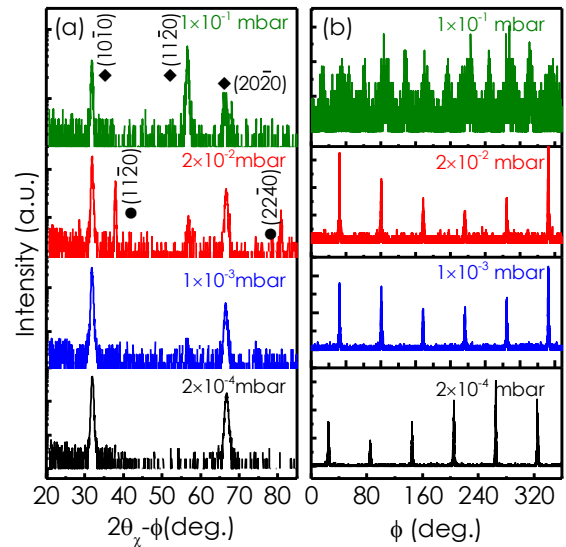

FIG S1.3 (a) In-plane $2\theta_\chi$-$\phi$ scan for the samples grown at different oxygen pressures. [symbols diamond and circle represent the peak from ZnO and sapphire, respectively] (b) wide angle $\phi$-scan profiles recorded for $(10\bar{1}0)$ ZnO reflection in these samples.

Electrical and luminesence properties

Figure S1.4 (a) shows the resistivity of the films as a function of growth pressure. The increase in resistivity of the films with $p_{O_2}$, suggests the formation of large number of compensating defects as the oxygen pressure increases. Hall measurement shows n-type conductivity in these films. This has to be noted that the Zinc vacancies act as acceptors in ZnO. The effect of increase of resistivity with $p_{O_2}$ can be attributed to the acceptor nature of the Zn-vacancies, which compensate the background n-type doping. Highest mobility and carrier concentration are found to be ~32 cm$^2$/V.s and ~$1.1 \times 10^{19}$ cm$^{-3}$ in the films grown below $p_{O_2} = 1 \times 10^{-3}$ mbar. Hall measurement on films grown with $p_{O_2} > 1 \times 10^{-3}$ mbar could not be carried out due to high resistive nature of the films. Figure S1.4(b) presents the normalized PL spectra obtained at room temperature for all the samples. PL peak normalized with respect to the near band-edge peak. At low oxygen pressure, the PL intensity is dominated by the near band-edge feature. Interestingly, as we increase the oxygen pressure, PL intensity corresponds to other defects gets enhanced. Two broad luminescence features at 2.1 and 2.52 eV could be seen at film grown at $p_{O_2} = 1 \times 10^{-1}$ mbar. The peak at 2.52 eV can be attributed to the transition between Zn-vacancy (0/-1) level to conduction band minimum, whereas the peak at 2.1 eV can be ascribed to the transition between Zn-vacancy (-/-2) defect level to valence band maximum[1]. It should be noted that the formation energy of Zn vacancy is lowest among all point defects when growth takes place in oxygen rich condition [2].

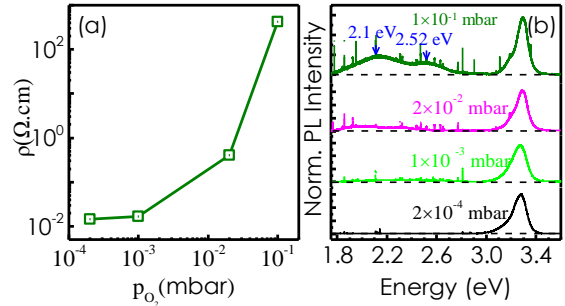

FIG. S1.4. (a) Resistivity (ρ) of the films as a function of oxygen pressure ($p_{O_2}$). (b) Room temperature photoluminescence (PL) spectra recorded for samples grown at different $p_{O_2}$.

## *S2.* Ohmic nature of the Ni/Au and Ti/Au contacts

Electrical measurements are performed on (001)NiO/(0001)ZnO heterojunction devices at room temperature. Ti(20 nm)/Au(100 nm) and Ni(20 nm)/Au(100 nm) contact pads are fabricated on ZnO and NiO sides of these devices, respectively. Fig. S2. 1 (a) and (b) shows the current-voltage profiles recorded between two Ni/Au metal contact pads on NiO sides and between two Ti/Au contact pads on ZnO sides of the NiO/ZnO heterostructures, respectively [device structure shown in the respective insets]. Clearly, ohmic or nearly ohmic nature of the Ni/Au contact pads is evident. Ti/Au contact pads on ZnO films also show ohmic nature.

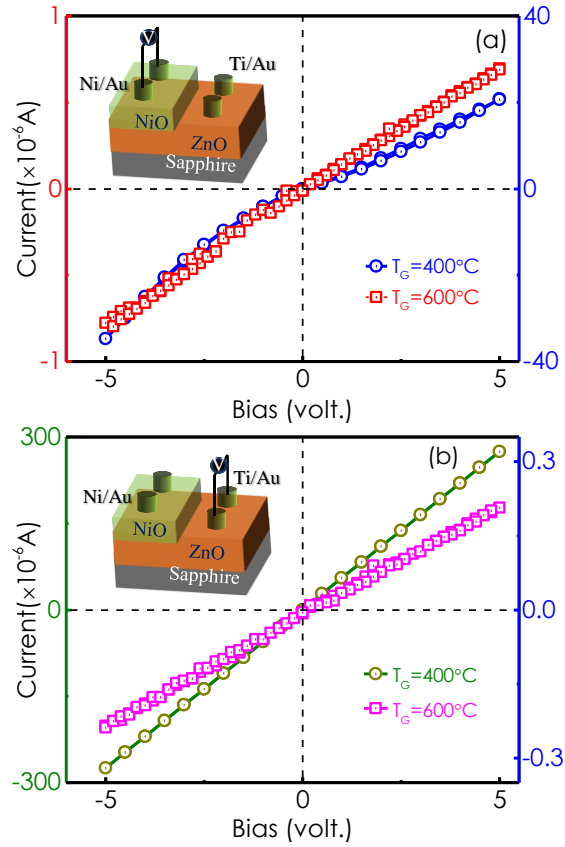

FIG. S2.1. Current-voltage (I-V) profiles recorded for the sample grown at different $T_G$ between (a) two Ni/Au contacts on the NiO side and (b) two Ti/Au contact pads on the ZnO side of NiO/ZnO heterostructures grown at different $T_G$. [Inset shows the schematic diagram of the devices].

## *S3.* **ZnO/NiO heterostructures**

These heterostructures are grown on c-sapphire substrates sapphire using pulsed laser deposition (PLD) technique. A NiO layer is first grown on the c-sapphire substrates at a certain growth temperature $T_G$ and oxygen pressure of $2 \times 10^{-2}$ mbar. Subsequently, ZnO layer is grown on top of the NiO film at a growth temperature of 500°C keeping the oxygen pressure unchanged. Several such heterostructures are grown, where only the growth temperature during NiO deposition ($T_G$) is varied from 200 to 500°C.

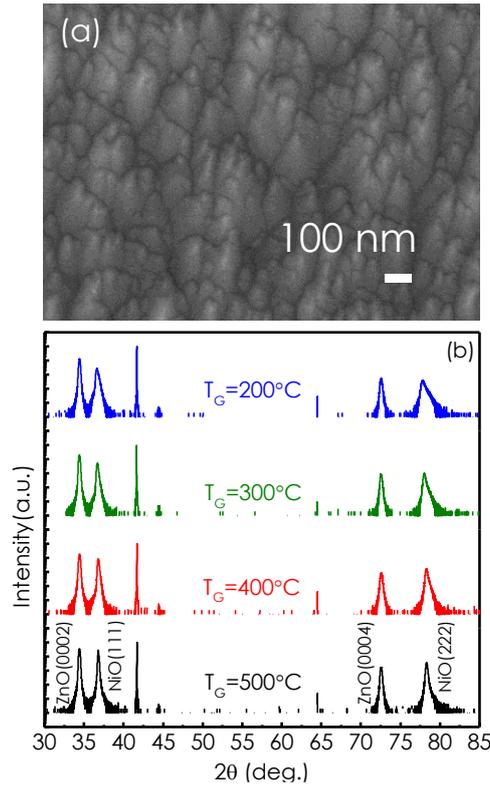

FIG. S3.1. (a) Surface SEM image of ZnO film grown on top of NiO template (b) $\omega$-$2\theta$ XRD scans for all ZnO samples for which the NiO template layers are grown at different temperatures ($T_G$).

Figure S3.1(a) represents the surface SEM image of ZnO film grown on top of NiO template. Clearly, the surface appears to have some columnar structures throughout the layer. Figure S3.1(b) represents $\omega$-$2\theta$ XRD scan of the films grown at different temperatures. Diffraction peaks associated with (0002) ZnO, (111) NiO and (0006) $Al_2O_3$ and their higher order reflections are the only peaks observed in the out of plane direction. This suggests the growth of c-oriented ZnO film on top of (111)NiO template. In-plane XRD study further reveals the epitaxial nature of the NiO films. However, ZnO grains are not found to show any orientation relationship in the in-plane directions. No rectifying behaviour could be seen in these heterostructure devices. This can be attributed to the columnar morphology of the ZnO films. It is plausible that the metal layers deposited on ZnO surface can propagate through the pores and make contacts with the underlying NiO layer.